# Brillouin Scattering Selection Rules in Polarization-Sensitive Photonic Resonators

Anne Rodriguez[1,*], Priya[1,*], Edson Cardozo de Oliveira[1], Luc Le Gratiet[1], Isabelle Sagnes[1], Martina Morassi[1], Aristide Lemaître[1], Florian Pastier[2], Loïc Lanco[3], Martin Esmann[1,+], Norberto Daniel Lanzillotti-Kimura[1,#]

[1] Université Paris-Saclay, CNRS, Centre de Nanosciences et de Nanotechnologies, 10 Boulevard Thomas Gobert, 91120 Palaiseau, France

[2] Quandela SAS, 10 Boulevard Thomas Gobert, 91120 Palaiseau, France

[3] Université Paris-Cité, CNRS, Centre de Nanosciences et de Nanotechnologies, 10 Boulevard Thomas Gobert, 91120 Palaiseau, France

[+] current affiliation: Institute of Physics, Carl von Ossietzky University, 26129 Oldenburg, Germany

[*] These authors contributed equally.　　[#] daniel.kimura@c2n.upsaclay.fr

**Abstract**

The selection rules governing spontaneous Brillouin scattering in crystalline solids are usually taken as intrinsic material properties, locking the relative polarization of excitation and signal in bulk. In this work, we independently manipulate these polarization states by means of optical resonances in elliptical micropillars and demonstrate a polarization-based filtering scheme for Brillouin spectroscopy in the 20-100 GHz range, important for telecom applications. This strong modification of selection rules using elliptical micropillars can be extended to any optical system with localized, polarization-sensitive modes, such as plasmonic resonators, photonic crystals, birefringent micro-, and nanostructures. Our polarization control protocol will thus find applications in the engineering of light-matter interactions in optomechanical, optoelectronic and quantum optics devices.

# Introduction

Brillouin scattering, inelastic scattering of light, is extensively used in material characterization, biological imaging, and optical and optoelectronic devices [1]. In Brillouin scattering processes, the selection rules formally constrain the energy, direction, and polarization of the scattered photons for a given input state. These selection rules in crystalline solids are usually taken as intrinsic material properties, locking the relative polarization of excitation and signal states [2]. For example, by exciting a zinc blende material such as GaAs along the (001) direction, the backscattered Brillouin signal preserves the polarization state of the excitation laser source. Using microstructures, these selection rules can be broken.

It has been shown that the wave vector selection rules of spontaneous Brillouin scattering can be modified in microstructures [3]. Also, the scattering cross-sections can be largely enhanced by means of micro- and nanostructures such as microcavities [3,4] and surfaces [5,6]. More recently, polarization control in stimulated Brillouin scattering has been reported in birefringent photonic crystal fibers, polarization maintaining fibers, and nanofibers [7–9]. However, the control of polarization in the spontaneous Brillouin scattering regime is seldom explored. In this work, we show that not only the wave vector selection rules and scattering cross-sections but also the polarization selection rules of spontaneous Brillouin scattering can be strongly modified in polarization-sensitive photonic resonators, such as elliptical micropillars [10,11], optical nanoantennas [12–14] and metasurfaces [15,16].

We introduce elliptical optical micropillar resonators to control polarization selection rules. Due to the anisotropy of the micropillar cross-section, it features two confined optical cavity eigenmodes with orthogonal polarizations [11,17,18] and non-degenerate energies. The energy separation of the modes can be controlled by the size and ellipticity of the pillar. For the particular case of a circular pillar, the modes are degenerate. The two orthogonal resonances can lead to an energy-dependent polarization rotation of light. This property of elliptical micropillars has already been employed for polarization-dependent emission of quantum-dot-based single-photon sources [10,19–22].
As we will show later, because of the energy difference resulting from the inelastic scattering process, the resonant Brillouin scattering emission undergoes a different rotation of polarization than the incident laser. By properly choosing the polarization and wavelength of the incident laser, we could reach a situation in which the Brillouin signal and the reflected laser are in orthogonal polarization states. In this way, background-free spontaneous Brillouin scattering spectra can be efficiently measured in a cross-polarization scheme.

# Results
## Polarization rotation by an elliptical micropillar resonator.

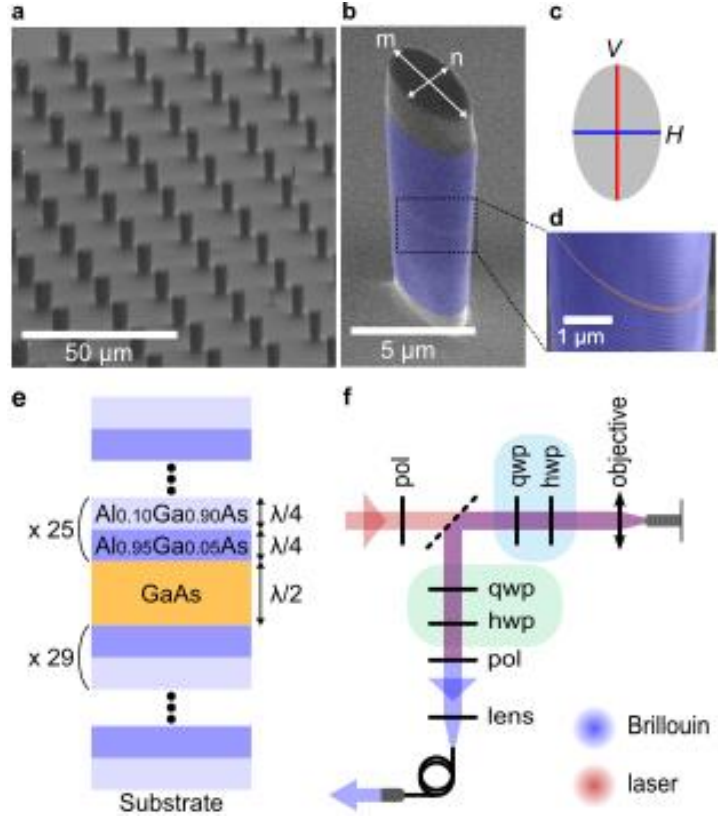

**Figure 1: Micropillar resonators and experimental setup. (a)** SEM image of an array of micropillars with various sizes and ellipticities. **(b)** SEM image of an elliptical pillar, *m,* and *n* are the major and minor axis lengths of the cross-section, respectively. The vertical structure consists of two GaAs/AlAs DBRs (blue) enclosing a resonant half-wavelength GaAs spacer layer (orange) **(c)** Schematic of the pillar cross-section. The two fundamental optical eigenmodes are of orthogonal linear polarizations *V* and *H*, polarized along the two axes of the cross-section. **(d)** Zoom in to the structure of a pillar. **(e)** Schematic of the vertical layer structure of the micropillar with two DBRs enclosing a resonant spacer. **(f)** Experimental scheme. The polarized excitation laser is focused to a spot of 2.2 µm on the sample with an objective lens of 0.7 NA. The Brillouin signal is collected through a single mode fiber before entering a double spectrometer in additive mode. *Pol* stands for polarizer and *qwp* (*hwp*) for a quarter (half) waveplate.

A micropillar cavity consists of two distributed Bragg reflectors (DBRs) enclosing a resonant spacer [23,24] (see Fig. 1). Figure 1b shows a scanning electron microscope (SEM) image of an individual micropillar, where the DBRs are highlighted in blue and the spacer in orange. An elliptical micropillar cavity has two fundamental optical eigenmodes |H⟩ and |V⟩ of orthogonal linear polarizations (horizontal/vertical, *H*/*V*). They are polarized along the minor/major axis of the elliptical cross-section, as shown in Fig. 1c, and have non-degenerate frequencies $\omega_{c,H}$ and $\omega_{c,V}$, respectively. In what follows, we establish the polarization states of the reflected laser and the Brillouin signal in this system.

Considering an incoming laser field of frequency $\omega_{in}$, we define the input polarization state as:

$$|\psi_{in}\rangle = \frac{1}{\sqrt{|b_{in,H}|^2 + |b_{in,V}|^2}} (b_{in,H}|H\rangle + b_{in,V}|V\rangle) \tag{1}$$

The associated intracavity field as:

$$|\psi_{cav}\rangle = \frac{1}{\sqrt{|a_H|^2 + |a_V|^2}} (a_H|H\rangle + a_V|V\rangle) \tag{2}$$

And the reflected field as:

$$|\psi_{refl}\rangle = \frac{1}{\sqrt{|b_{out,H}|^2 + |b_{out,V}|^2}} (b_{out,H}|H\rangle + b_{out,V}|V\rangle) \tag{3}$$

In Eqns. (1-3), $b_{in,H/V}$, $b_{out,H/V}$ and $a_{H/V}$ are the polarization amplitudes of the incoming, reflected, and intracavity laser fields, respectively. In Eq. (2), $a_{H/V} \propto b_{in,H/V} \frac{1}{1 - 2i\frac{\omega_{in} - \omega_{c,H/V}}{\kappa_{H/V}}}$ [25,26] with $\kappa$ the cavity damping rate.

The reflected fields in Eq. (3) are obtained using the standard input-output equations $b_{out,H/V} = b_{in,H/V} + \sqrt{\kappa_t} \times a_{H/V}$, with $\kappa_t$ the polarization-dependent top DBR leakage rate. $b_{out}$ represents the reflected field as an interference between the input light directly reflected by the top DBR and the light emerging from the cavity.

The Brillouin scattering field inside the GaAs cavity spacer has the same polarization amplitudes as the intracavity polarization state $|\psi_{cav}\rangle$ of the laser but at a different frequency $\omega_B$. The polarization state of Brillouin scattering $|\psi_B\rangle$ outside the cavity is then given by:

$$|\psi_B\rangle = \frac{1}{\sqrt{|b_{B,H}|^2 + |b_{B,V}|^2}} (b_{B,H}|H\rangle + b_{B,V}|V\rangle), \tag{4}$$

where $b_{B,H/V} \propto \frac{1}{1 - 2i\frac{\omega_B - \omega_{c,H/V}}{\kappa_{H/V}}} \times a_{H/V}$. Here, $b_{B,H/V}$ is proportional to the product of two terms: the first term depends on the frequency of the Brillouin signal, and the second term depends on the incoming field frequency and polarization through the intracavity amplitude $a_{H/V}$. We find that the polarization state of the Brillouin signal can be controlled by the micropillar geometry and is not dictated only by the polarization selection rules. Moreover, the reflected laser and the Brillouin signal polarization states experience different degrees of polarization rotation and can hence be discriminated by polarization filtering.

For the experimental demonstration of these phenomena, we study a sample consisting of arrays of GaAlAs micropillars with various sizes and ellipticities (Fig. 1a). Vertically, the pillars consist of two ($\lambda/4$ / $\lambda/4$) GaAs/AlAs DBRs enclosing a resonant $\lambda/2$ GaAs spacer layer. They act as an optical resonator for near-infrared photons and as an acoustic resonator for longitudinal acoustic phonons around 18 GHz [27–30]. We define the micropillar ellipticity as $e = \sqrt{m/n} - 1$ [11], where $m$ and $n$ are the major and minor axis lengths of the elliptical cross-section, see Fig. 1b, c.

Figure 1d presents the experimental spectroscopy setup to measure reflectivity and Brillouin scattering. The polarization of the incident laser and the collected signal can be controlled independently.

We measure the polarization-dependent optical reflectivity (Fig. 2) for an elliptical micropillar of ellipticity $e = 0.41$ with $m = 4\ \mu m$. We prepare an incident laser with diagonal polarization $|\psi_{in}\rangle = \frac{1}{\sqrt{2}}(|H\rangle + |V\rangle) = |D\rangle$ and detect $|\langle\psi_{det}|\psi_{refl}\rangle|^2$ with $|\psi_{det}\rangle = |H\rangle, |V\rangle$. By scanning the laser frequency, we observe two well-defined polarization-dependent optical modes (panel a, top). The blue (red) mode corresponds to $H(V)$ polarization, associated with the minor (major) axis of the pillar cross-section, respectively. Using the definitions in Eqs. (1-3), the polarization-dependent complex reflection coefficients $r_V$ and $r_H$ are

$$r_{H/V} = \frac{b_{out,H/V}}{b_{in,H/V}} = 1 + \sqrt{\kappa_t}\frac{1}{1 - 2i\frac{\omega_{in} - \omega_{c,H/V}}{\kappa_{H/V}}} \qquad (4)$$

The cavity damping $\kappa_{H/V}$ includes sidewall losses and leakage through the top and bottom DBRs [26]. In panel a (bottom), we measure the antidiagonal $|A\rangle$ component of the reflected signal, i.e. $|\psi_{det}\rangle = |\overline{\psi_{in}}\rangle = |A\rangle$ collecting light in a cross-polarization scheme. Effectively, this signal represents laser photons, whose linear polarization state has been rotated from $|D\rangle$ to $|A\rangle$ upon reflection. We observe two maxima that are roughly localized at the spectral position of the eigenmodes of the elliptical micropillar. Outside the resonances, the rotated signal is practically zero. We simulate these rotated signals using the Jones matrices formalism (panel b) with an excellent agreement (see Methods section). To convey the full information of the reflected signals, we plot the simulated reflection state on a Poincaré sphere in panel c. As a function of wavelength, the reflected state undergoes a trajectory that spans the full sphere.

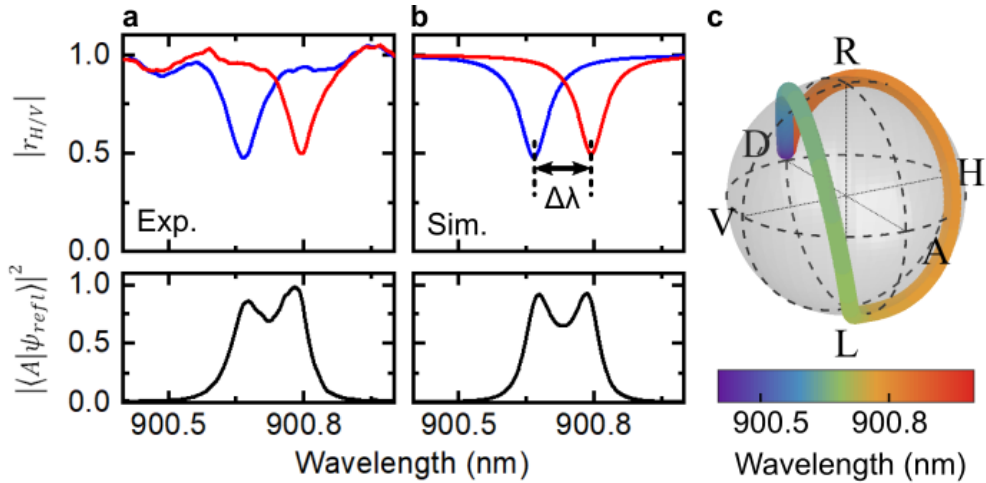

**Figure 2: Polarization rotation by an elliptical micropillar resonator. a** Experimental reflectivity spectra of an elliptical micropillar of ellipticity $e = 0.41$ with $m = 4\ \mu m$. The blue (red) spectrum is measured with a linear polarization aligned with the minor (major) axis of the elliptical pillar cross-section. Two clear optical modes are observed at different central wavelengths. Bottom panel: Spectrum of the reflected laser with rotation of polarization.

The incident laser is polarized along *D*, while the detection is along *A*. **b** Calculated reflectivity spectra of the two optical modes presented in panel a. Bottom panel: Calculated spectrum of the reflected laser with rotation of polarization. $\Delta\lambda = 0.127$ nm is the splitting between the two linearly polarized eigenmodes. **c** Poincaré sphere displaying the calculated wavelength-dependent polarization state $|\psi_{refl}\rangle$ of the reflected laser.

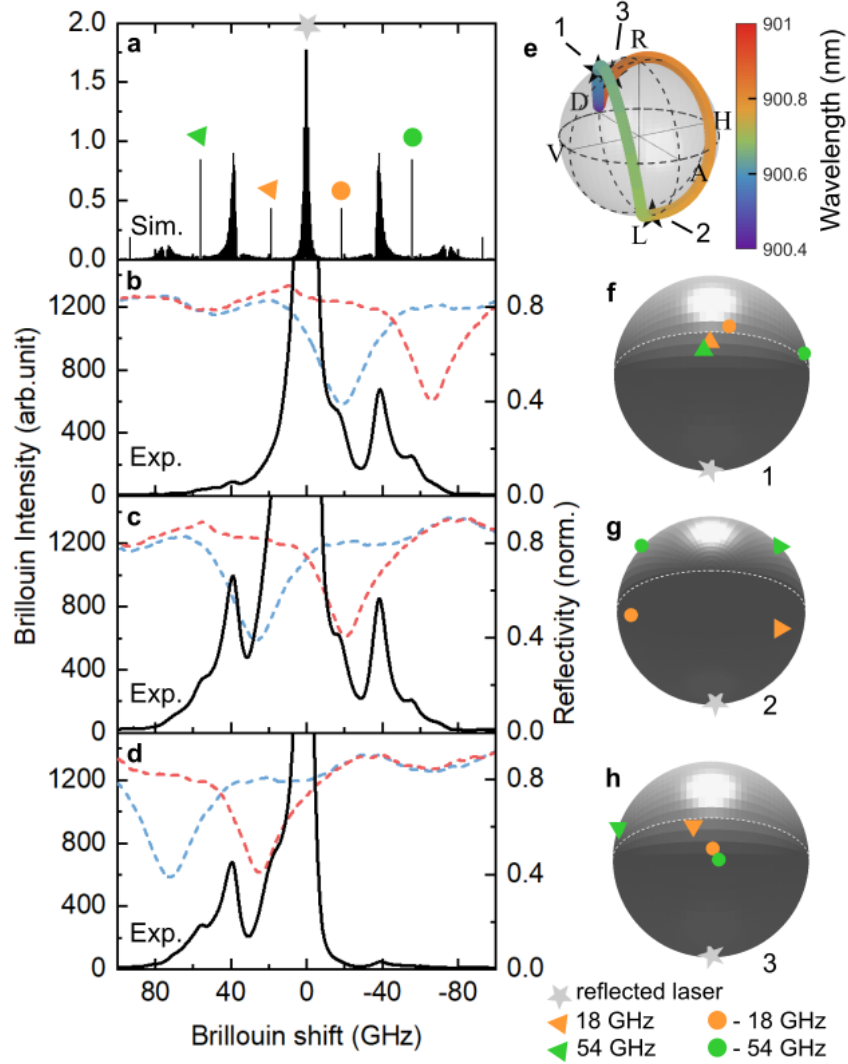

**Figure 3: Polarization rotation-enabled optical filtering. a** Simulated Brillouin spectrum for a planar cavity with the multilayer structure shown in Fig. 1e. Experimental Brillouin spectra acquired with the excitation laser blue-detuned from the optical modes (**b**), between the optical modes (**c**) and red-detuned from the optical modes (**d**). The peak at $\pm 18$ GHz corresponds to the fundamental confined acoustic mode of the resonator. The peak at $\sim\pm 37$ GHz corresponds to bulk Brillouin scattering from the substrate. The peak at $\pm 54$ GHz corresponds to the third harmonic of the confined mode. **e** Simulated polarization states of the reflected laser for the wavelengths used in panels b-d indicated with stars on the Poincaré sphere. **f, g, h** Simulated polarization of the reflected laser and Brillouin signals (b,c,d, respectively) plotted on Poincaré spheres rotated such that the reflected laser state is localized at the South pole. The excursion of the Brillouin states from the South pole is a direct indication of strongly modified polarization selection rules enabling a polarization filtering protocol.

**Polarization rotation-enabled optical filtering.**

We exploit the difference in rotation of polarization between the reflected laser field and Brillouin scattering for a polarization filtering protocol. Detecting the Brillouin scattered signal requires filtering out the laser light. The difference in rotation of polarization implies that $|\psi_{in}\rangle \neq |\psi_{refl}\rangle \neq |\psi_B\rangle$. That is, we alter the backscattering Brillouin selection rule in bulk GaAs ($|\psi_{in}\rangle = |\psi_{refl}\rangle = |\psi_B\rangle$) by engineering the optical states in a cavity. We achieve the filtering by detecting the Brillouin signal in a cross-polarization geometry ensuring that $\langle\psi_{det}|\psi_{refl}\rangle = 0$. That is:

$$|\psi_{det}\rangle = |\overline{\psi_{refl}}\rangle = \frac{1}{\sqrt{|b_{out,H}|^2 + |b_{out,V}|^2}}(b_{out,V}{}^*|H\rangle - b_{out,H}{}^*|V\rangle) \tag{7}$$

Note that here the cross-polarization condition is wavelength-dependent, due to the wavelength-dependence of $|\psi_{refl}\rangle$ as shown in Fig. 2.

In Fig. 3a we show a simulated Brillouin spectrum (see Methods section) of the vertical microcavity structure (see Fig. 1 e). We observe peaks at $\pm 18$ GHz, $\pm 54$ GHz, and $\pm 90$ GHz corresponding to the Stokes and anti-Stokes harmonics of the acoustic modes confined in the cavity [27,31,32]. The peaks at $\sim\pm 37$ GHz correspond to bulk Brillouin scattering from the substrate. Panels b-d correspond to experimental Brillouin spectra measured at excitation laser wavelengths of 900.615 nm, 900.737 nm, and 900.860 nm. For reference, the polarization-dependent optical reflectivity is included in each panel with dashed lines. Panel e retakes the information plotted in Fig. 2c, where numbered stars indicate the polarization state of the reflected laser for each of the measured cases. Panels f-h show on the same sphere the simulated polarization states of the reflected laser, $|\psi_{refl}\rangle$, and the Brillouin scattered signals $|\psi_B\rangle$. The polarization states are simulated using Jones matrices and considering Brillouin scattering as a source inside the cavity spacer. For clarity reasons, we reorient the Poincaré sphere such that $|\psi_{refl}\rangle$ is systematically located at the South pole of the sphere, while $|\psi_{det}\rangle$ is located at the North pole.

Figure 3b displays a Brillouin spectrum obtained with the excitation laser blue-detuned from the cavity modes, at a wavelength of 900.615 nm. The observed Stokes Brillouin spectrum shows that the fundamental mode at -18 GHz is resonant with the *H*-polarized cavity mode and the third harmonic, at -54 GHz, is coupled to the *V*-polarized cavity mode. The anti-Stokes modes appear with much less intensity in the measurement. Panel f shows that the four relevant signals experience a rotation of polarization that makes them detectable. All of them are away from the South pole, representing the reflected laser, which we filter out. While the polarization projection is similar in the four cases, the Stokes components are enhanced by the presence of the optical cavity modes.

Figure 3c shows a Brillouin spectrum measured with the laser tuned between the optical cavity modes, at a wavelength of 900.737 nm. As shown in the Poincaré sphere (panel g), both Stokes and anti-Stokes scattering modes at $\pm 18$ and $\pm 54$ GHz have polarizations away from the South pole, with the modes at $\pm 54$ GHz closer to the North pole, implying a better polarization filtering condition for these modes. In contrast to panel b, both the Stokes and anti-Stokes modes are enhanced by the coupling to the optical cavity modes.

To selectively measure the anti-Stokes components in a configuration where they are enabled by the cross-polarization filtering and enhanced by the coupling to the optical cavity modes, we red-detune the laser. Figure 3d shows a corresponding spectrum obtained with the laser at 900.860 nm. The observed anti-Stokes Brillouin spectrum shows the mode at +18 GHz which is coupled to the *V*-polarized optical cavity, and the mode at +54 GHz coupled to the *H*-polarized cavity mode. The polarization states on the Poincaré sphere (panel h) are mirror images of the states in panel f, i.e. all the modes are away from the South pole.

By exploiting polarization rotation, we demonstrate that in an elliptical micropillar it is possible to selectively measure different bands of the Brillouin spectrum by properly choosing the excitation wavelength and polarization states. The optimum configuration for cross-polarized filtering would be obtained when the reflected laser and the Brillouin signal are orthogonal, i.e. in opposite poles of the Poincaré sphere $\langle \overline{\psi_{refl}} | \psi_B \rangle = 1$.

So far, we have discussed the excitation conditions as a means to optimize the measurement efficiency. In Fig. 4, we theoretically analyze the effects of changing the ellipticity of the micropillar on the scattering selection rules. Fig. 4a shows the polarization state of the Brillouin signal as a function of the micropillar ellipticity while keeping the excitation wavelength centered between the two optical cavity modes and the input polarization along *D*. The trajectory of the Brillouin signal shows that the relation between the incident laser and signal polarization states is strongly modified, breaking the intrinsic selection rules. Fig. 4b shows the reflected laser polarization state under the same excitation conditions. We observe that the resulting state is always different from the corresponding Brillouin states $\langle \psi_{refl} | \psi_B \rangle \neq 1$, meaning that cross-polarized detection is always possible for non-degenerate polarization-dependent optical modes. The accessible phonon frequency band is determined by the separation of the optical cavity modes. In elliptical micropillars, the separation typically reaches up to 10 meV (2.5 THz) [10,11].

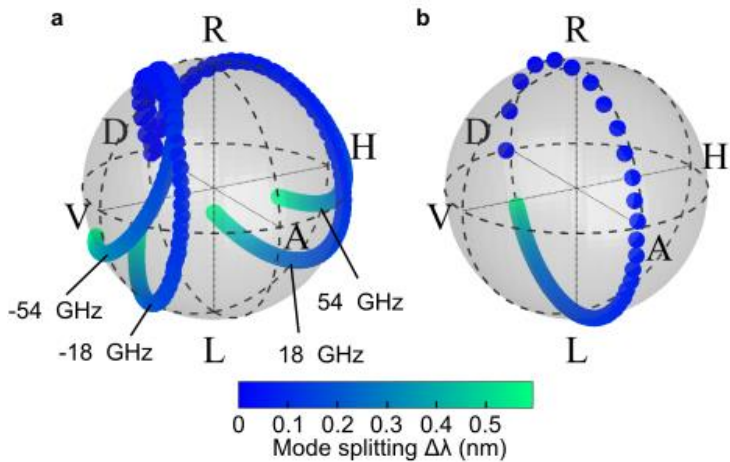

**Figure 4: Effect of the micropillar ellipticity on the scattering selection rules. a** Brillouin polarization states for the fundamental and third harmonic of the confined acoustic mode as a function of the mode splitting Δλ between the two optical modes in elliptical micropillars. We consider the particular case of an excitation laser with diagonal polarization and tuned between the two optical modes. For zero separation, the Brillouin polarization states coincide

with the excitation laser. By increasing the separation, the modes describe frequency-dependent trajectories that span the Poincaré sphere. **b** Reflected laser polarization state as a function of the separation between the two optical modes in elliptical micropillars with the same excitation conditions as in a. The described trajectory evolves along a meridian of the Poincaré sphere. The difference in polarization states observed between panels a and b enables an efficient polarization filtering protocol.

## Conclusion

We have theoretically proposed and experimentally demonstrated a strategy to independently manipulate the polarization state of a Brillouin scattered signal and the excitation laser. This strong modification of the polarization selection rules is achieved in elliptical micropillar cavities presenting polarization-sensitive optical cavity modes. Due to the wavelength-dependent birefringence of the elliptical pillar, the reflected laser beam and the Brillouin scattered signal encounter different rotations of polarization, enabling a cross-polarized filtering scheme. Using the laser wavelength as a measurement parameter, we selectively access different frequency bands in the Stokes and anti-Stokes Brillouin scattering spectra. In the studied case, the cavity mode lifetimes, spectral separation determined via the ellipticity, and laser wavelength and polarization define the parameter space to maximize cavity-enhanced signals under optimal polarization-filtering conditions. The same working principle applies to any photonic system with localized, polarization-sensitive modes, such as plasmonic resonators, photonic crystals, and birefringent micro- and nanostructures.

The presented technique is particularly important for studying phonons with frequencies between 1 GHz and 1 THz, relevant for thermal transport and telecommunication applications. In this range, standard Raman spectroscopy techniques lack enough resolution, and standard Brillouin scattering techniques lack versatility [33,34]. Our polarization control protocol will thus find applications in the engineering of light-matter interactions in optomechanical, optoelectronic and quantum optics devices [10,35,36].

# Methods

**Fabrication details.** The sample under study is grown on a (001)-oriented GaAs substrate by molecular-beam epitaxy. It consists of an acousto-optical microcavity with two DBRs enclosing a resonant spacer with an optical path length of $\lambda/2$ at a resonance wavelength of around $\lambda \sim 900$ nm. The top (bottom) optical DBR is formed by 25 (29) periods of $Ga_{0.9}Al_{0.1}As/Ga_{0.05}Al_{0.95}As$ bilayers ($\lambda/4 / \lambda/4$) (fig.1e). From this planar acousto-optical cavity, micropillars of various sizes and ellipticities are fabricated by optical lithography and inductively coupled plasma etching (fig.1a). The micropillars confine an optical mode with typical Q-factors of 11 000. This structure confines acoustic phonons at around 18 GHz [28,29,37].

**Cross-polarized experimental scheme.** Figure 1(f) shows a schematic of the Brillouin spectroscopy setup implemented using polarization as a filtering technique in a backscattering configuration. A collimated laser beam from a tunable continuous wave (CW) Ti:Sapphire laser (M2 SolsTis) is used as an excitation source. The incident beam polarization is initialized with a polarizer. A second polarizer is placed in the collection path. A quarter-waveplate and a half-waveplate (qwp and hwp) control the incident polarization state on the sample $|\psi_{in}\rangle$. We set the incident beam into diagonal polarization. The incident laser beam is focused on the sample with a spot diameter of approximately 2.2 µm using a NA = 0.7 objective lens. We collect the reflected signal through the same objective and waveplates. A second set of waveplates placed in front of the polarizer in the collection path allows us to choose the polarization basis for the collection, while the second polarizer acts as an analyzer. For Fig. 2 we measure the polarization-dependent reflectivity by analyzing it on the *H/V* basis. To measure the polarization rotation, we excite with *D* polarization and collect in *A*.

For Fig. 3, we use an incident laser power of 50 µW with typical integration times of 0.1 s. The Brillouin signal emerging from the sample is collected through the same objective and waveplates. The second set of waveplates in front of the polarizer in the collection path allows us to extinguish the laser before the remainder of the light is coupled into a single mode fiber. The transmission of the polarizer at the collection is 86% at the wavelength of interest ~900 nm. The extinction ratio of the laser is 45:1 when Stokes and anti-Stokes are coupled to the *V* and *H* optical modes and 78:1 when only Stokes is coupled to a cavity mode, measured before the collection fiber. The use of a single mode fiber increases the purity of the signal through spatial filtering [33]. The Brillouin signal is finally analyzed with a double spectrometer operating in additive mode.

For the wavelength-dependent measurements in Fig. 3, we always reconfigure the waveplates in the collection path such that we project along $|\overline{\psi_{refl}}\rangle$.

**Simulations** For the simulation of the Brillouin spectrum in Fig. 3a, we assume a purely photoelastic interaction [31]. For the numerical implementation, we use the transfer matrix method with nominal material properties and assume a planar structure. The spectrum was convoluted with a Gaussian resulting in a resolution of 0.025 GHz.

For the polarization-dependent studies, we based our simulation on the Jones matrices formalism. We fit the reflectivity contrast, resonance wavelength, and linewidth of the polarization-dependent eigenmodes of the elliptical micropillar using a Lorentzian model (Fig. 2b top). We then use the resulting parameters as input for the Jones matrices to compute the simulated results shown in Fig. 2b, bottom, Fig. 2c, and Fig. 3e-h. For Fig. 4 we take the same linewidth and reflectivity contrast as inputs.

**Acknowledgments**

The authors gratefully acknowledge P. Senellart for fruitful discussions, I. Ahmane for experimental support at an early stage of the project, and S. Ravets for the dielectric deposition. The authors acknowledge funding from the European Research Council Starting Grant No.715939, Nanophennec. This work was supported by the European Commission in the form of the H2020 FET Proactive project TOCHA (No. 824140). The authors acknowledge funding from the French RENATECH network and through a public grant overseen by the ANR as part of the Investissements d'Avenir" program (Labex NanoSaclay Grant No.ANR-10-LABX-0035). M.E. acknowledges funding from the Deutsche Forschungsgemeinschaft (DFG, German Research Foundation) Project 401390650.